\documentclass[osajnl,twocolumn,showpacs,superscriptaddress,10pt]{revtex4-1}
\usepackage[latin1]{inputenc}
\usepackage{graphicx}

\begin{document}

\title{Vibration of low amplitude imaged in amplitude and phase by sideband versus carrier correlation digital  holography}

\author{N. Verrier$^{1,2}$,  L. Alloul$^1$ and M. Gross}

\affiliation{Laboratoire Charles Coulomb  - UMR 5221 CNRS-UM2
Université Montpellier II Bat 11.
Place Eugène Bataillon 34095 Montpellier\\
$^2$Laboratoire Hubert Curien UMR 5516 CNRS-Universit\'{e} Jean Monnet 18 Rue du
Professeur Beno\^{\i}t Lauras 42000 Saint-Etienne}



\begin{abstract}
Sideband  holography can be used to get fields images ($E_0$ and $E_1$) of a vibrating object for both the carrier ($E_0$) and the sideband ($E_1$) frequency with respect to vibration. We propose here to record  $E_0$ and  $E_1$ sequentially, and to image the correlation $E_1 E_0^*$. We show that this correlation is insensitive the phase related to the object roughness and directly reflect the phase of the mechanical motion. The signal to noise can  be improved by averaging  the correlation over neighbor pixel.  Experimental validation is made with vibrating cube of wood and with a clarinet reed. At 2 kHz, vibrations of amplitude down to 0.01 nm are detected.
\end{abstract}


\maketitle

\textbf{Citation: }
N. Verrier, L. Alloul, and M. Gross, Opt. Lett. 40, 411-414 (2015) 

\url{http://dx.doi.org/10.1364/OL.40.000411}
\bigskip

There is a big demand for full field vibration measurements, in particular in industry. Different
holographic techniques are able to image and analyze such vibrations. Double pulse
\cite{pedrini1995digital,pedrini1997digital} or multi pulse
holography \cite{pedrini1998transient}  records several holograms with time separation in the 1...1000 $\mu$s range, getting the instantaneous velocity from the phase difference. If the vibration frequency is not too high, one can also directly track the vibration of the object with fast CMOS cameras \cite{pedrini2006high,fu2007vibration}. The  analysis of the
motion can be done by phase difference or by Fourier analysis in the time domain. For  periodic vibration motions, measurements can be done with slow camera. Indeed,  an harmonically vibrating object illuminated by a laser yields  alternate dark and bright fringes \cite{powell1965interferometric}, that can be analyzed by time averaged holography \cite{picart2003time}.   Although the time averaged method gives a way to determine the amplitude of vibration \cite{picart2005some} quantitative measurement remains quite difficult for low and high vibration amplitudes.

We have developed heterodyne holography \cite{le2000numerical,le2001synthetic}, which is a variant of phase shifting holography, in which the frequency, phase and amplitude of both reference and signal beam are  controlled by acousto optic modulators (AOM).  Heterodyne holography is thus extremely versatile.
By shifting the frequency of the local oscillator $\omega_{LO}$ with respect to illumination $\omega_0$, it is for example possible  to detect the holographic signal at a frequency $\omega$ different  than illumination $\omega_0$. This ability is extremely useful to analyze vibration, since  heterodyne holography can detect selectively the signal that is scattered by object on vibration sideband of frequency   $\omega_m=\omega_0+ m \omega_A$, where $\omega_A$ is the vibration frequency and  $m$ and integer index.
%
%

As was reported by Ueda et al, \cite{ueda1976signal} the detection of the sideband $ m$=1 is advantageous when the vibration amplitude is small. Nanometric
vibration amplitude measurements were achieved with sideband digital  holography on the $m=1$ sideband
\cite{psota2012measurement}, and comparison with  single point  laser interferometry has been  made \cite{psota2012comparison}.
Verrier et al.  \cite{verrier2013absolute} have shown that one can simultaneously measure $E_0$ and $E_1$ by using a local oscillator with two frequency components. One can thus  infer the mechanical phase of the vibration \cite{bruno2014phase}. However, Bruno et al  \cite{bruno2014holographic} shows that this simultaneous detection of $E_0$ and $E_1$ can lead to cross talk, and to a loss of detection sensitivity, which becomes  annoying when the vibration amplitude is small.

In this letter, we show that simultaneous detection of $E_0$ and $E_1$ is not necessary, and that equivalent or superior performances can be obtained by detecting $E_0$ and $E_1$ sequentially. Indeed, the  cross talk effects seen by Bruno et al. \cite{bruno2014holographic} disappear in that case. We also show that the random phase variations caused by the roughness of the object can be eliminated by calculating the correlation $E_1 E_0^*$. It is then possible to increase the signal to noise ratio (SNR) by averaging correlation over neighboring pixels.
The sequential measurement of $E_0$ and $E_1$, and the calculation of the correlation $E_1 E_0^*$, make possible to image the vibration ``full field'', and to measure quantitatively its amplitude and phase. Maximum sensitivity is achieved by focusing the illumination in the studied  point and by averaging the correlation in that region. Finally, we prove that the sensitivity is limited by a sideband signal of one photo-electron per demodulated image sequence. The device and method were validated experimentally by studying a cube of wood  vibrating at $\simeq 20$ kHz, and a clarinet reed at $\simeq$ 2 kHz.
Measurement sensitivity of 0.01 nm for a vibration at 2 kHz, comparable to the sensitivity obtained by Bruno et al. \cite{bruno2014holographic} at 40 kHz, is demonstrated.
%
%
%


Consider an   object    illuminated by a laser at frequency $\omega_0$ that vibrate at frequency $\omega_A$ with an out of plane vibration amplitude $z_{max}$. The out of plane coordinate is $z(t)=z_{max} \textrm{sin} (\omega_A t)$. The field scattered by the object is ${\cal E}= E(t) e^{i\omega_0 t}+\textrm{ c.c.} $, where c.c. is the complex conjugate and $E(t)$  the field complex amplitude. In reflection geometry, we have
$$
E(t)= E_{wo} e^{ |\Phi| \textrm{sin} (\omega_A t+ \arg \Phi)}
$$
%
where $E_{wo}$ is the complex field without movement, and    $\Phi$ is a complex quantity that describes the phase modulation. The phase of this modulation is $\arg \Phi$, while  the amplitude is $|\Phi|=4\pi z_{max}/\lambda$.
Because of the Jacobi-Anger expansion, we have
$
E(t)= E_{wo} ~\sum_m J_m(|\Phi|)~e^{j m(\omega_A t+\arg \Phi) }
$
%
%
where  $J_m$ is the mth-order Bessel function of the first kind. The scattered field ${\cal E}$ is then the sum of  fields components ${\cal E}_m= E_m e^{i\omega_m t}+\textrm{ c.c.}  $ of frequency $\omega_m = \omega_0 + m \omega_A$, where $m$ is the sideband index ($m=0$ for the carrier) with $ E_m =E_{wo}~ J_m(|\Phi|)~e^{j m\arg \Phi} $.
When the vibration amplitude $\Phi$ becomes small, the energy  within sidebands  decrease very rapidly with the sideband indexes $m$, and  one has only to consider the carrier $m=0$, and the  first sideband $m=\pm1$.
We have thus:
\begin{eqnarray}\label{Eq_E_0_lim0}
E_0(\Phi) &=&  E_{wo}~J_0(|\Phi|)
\\
 \nonumber  E_{1}(\Phi) &=& E_{wo}~ J_{1}(|\Phi|) ~e^{j \arg \Phi} 
\end{eqnarray}
Note  that time averaged holography \cite{picart2003time} that detects the carrier field $E_0$  is  not efficient in detecting small amplitude vibration $|\Phi|$, because  $E_0$  varies quadratically  with $|\Phi|$. On the other hand,  sideband holography that is able to detect selectively the  sideband field $E_{1}$ is  much more sensitive, because  $E_{1}$  varies with  linearly with $|\Phi|$.
\begin{figure}[]
\begin{center}
\includegraphics[width=8 cm]{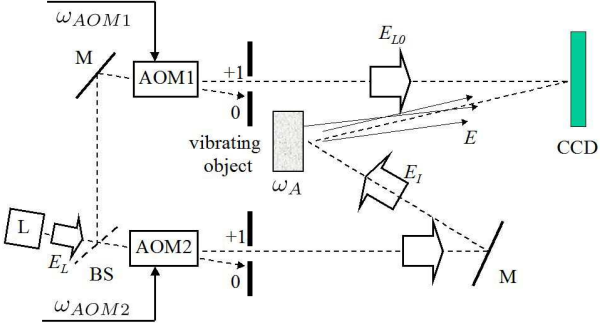}
 \caption{Heterodyne holography setup applied to analyse vibration. L: main laser; AOM1, AOM2: acousto-optic modulators; M: mirror; BS: beam splitter; CCD:  camera.}\label{Fig_fig_setup_reed}
\end{center}
\end{figure}

One must notice that the  field  scattered by the sample without vibration $E_{wo}$ depends strongly on the $x,y$ position. In a typical experiment, the sample rugosity is such that this field  is a fully developed speckle. The reconstructed  fields $E_{0}$ and $E_{1}$  are  thus random in phase, from one pixel $(x,y)$ to the next  $(x+1,y)$. One cannot thus simply extract the phase of the  vibration from a measurement made on a single sideband.
To remove the pixel to pixel  random phase of $E_{wo}$, we propose to record the hologram  successively on the carrier $m=0$ and the sideband $m=1$, to reconstruct the corresponding field image of the object $E_{0}(x,y)$ and $E_{1}(x,y)$, and to   calculate and image the correlation $E_1 E^*_0$, since this correlation do not involves  $E_{wo}$, but  $|E_{wo}|^2$, which is real and has no phase. Indeed, we have:
\begin{eqnarray}\label{Eq_E1_E0}
 E_1~E^*_0 &=& |E_{wo}|^2~ J_1(|\Phi|) ~J_0(|\Phi|) ~e^{j \arg \Phi}
\end{eqnarray}
For small vibration amplitude   ($|\Phi| \ll 1$), correlation simplifies to  $ E_1~E^*_0 \simeq |E_{wo}|^2 \Phi/2$. Correlation $E_1~E^*_0$  is a powerful tool since gives directly the phase of mechanical motion $\arg \Phi$. Nevertheless, problems can be encountered when the signal $|E_{wo}|^2$ scattered without vibration vanish.

%
%


Figure  \ref{Fig_fig_setup_reed} shows the holographic experimental setup used to measure successively $E_0$ and $E_1$ in order to get $E_1 E^*_0$. The main laser L is a Sanyo DL-7140-201
diode laser ($\lambda$=785 nm, 50 mW ).
It is split into an illumination beam (frequency $\omega_I$, complex
field $E_I$), and in a LO beam ($\omega_{LO}, E_{LO}$). The illumination light scattered by the object interferes with the reference beam on the camera (Lumenera 2-2: $1616 \times 1216$  pixels of $ 4.4\times  4.4 \mu$m) whose  frame rate  is $\omega_{CCD}$=10 Hz. To simplify further Fourier transform calculations, the  $1616 \times 1216$ measured matrix is truncated to $1024 \times 1024$.

The illumination and LO beam frequencies $\omega_{LO}$ and $\omega_I$ are  tuned by using two acousto-optic
modulators  AOM1 and AOM2 (Bragg cells), and we have
$\omega_{LO} =\omega_{L}+\omega_{AOM1}$ and $\omega_{I} =\omega_{L}+\omega_{AOM2}$, where $\omega_{AOM1/2}\simeq$80 MHz are the frequencies of RF signals that drive the AOMs.
The RF signals  are tuned to have $\omega_{LO}-\omega_I=\omega_{CCD}/4$  to get $E_0$, and to have  $\omega_{LO}-\omega_I= \omega_A+\omega_{CCD}/4$ to get $E_1$.
Successive sequences of $n_{max}=128$ camera frames (i.e. $I_0, I_1 ...I_{127}$) are recorded by tuning the RF signals first on the carrier ($E_0$), then on the sideband ($E_1$). The carrier and sideband complex hologram $H$ are obtained from these sequences by 4 phase demodulation with $n_{max}$ frames:
\begin{equation}\label{EQ_H_n_max}
  H(x,y)=\sum_{n=0}^{n=n_{max}-1}  j^n I_n(x,y)
\end{equation}
The fields images of the object $E_0(x,y)$ and  $E_1(x,y)$ are then reconstructed from $H$ by the Schnars et. al \cite{schnars1994direct} method that involves 1 Fourier transformation. The correlation $E_1 E^*_0$ is then calculated.

\begin{figure}
\begin{center}
  \includegraphics[width=8 cm]{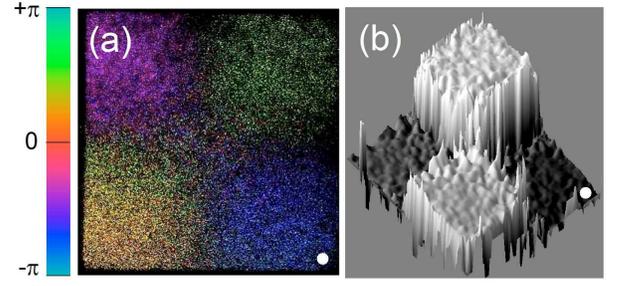}
 \caption{(a,b)  Reconstructed images of a cube of wood vibrating at $\omega_A/2\pi$= 21.43 kHz. (a) $E_1 E^*_0$ correlation image: brightness is  amplitude (i.e. $|E_1 E^*_0|$), color is phase (i.e. $\arg E_1 E^*_0$). (b) 3D display of  the phase $\arg E_1 E^*_0$.} \label{Fig_fig_vibrating_cube}
\end{center}
\end{figure}

Figure \ref{Fig_fig_vibrating_cube} (a) shows  the reconstructed correlation images of a cube of wood (2 cm$\times$2 cm) vibrating at its resonance frequency $\omega_A/2\pi$= 21.43 kHz. Brightness is the correlation  amplitude (i.e. $|E_1 E^*_0|$) and  color  the  correlation phase (i.e. $\arg E_1 E^*_0$). As seen, neighbor point have the same phase (same color). In order to get better SNR for the phase, the complex correlation signal $E_1 E^*_0$ is averaged over neighbor $x,y$ points by using a 2D Gaussian blur filter of radius 4 pixels. Figure \ref{Fig_fig_vibrating_cube} (b) displays the phase of  the averaged correlation $E_1 E^*_0$  in 3D. As seen,  the opposite corners (upper left and bottom right for example) vibrate in phase, while the neighbor corners (upper left and upper right for example) vibrate in phase opposition. Note that the cube is excited in one of its corner by a needle. This may explain why the opposite corners are not perfectly in phase in Fig. \ref{Fig_fig_vibrating_cube} (c).
%
%

We can go further and use $E_0$ and $E_1$ to calculate the vibration complex amplitude $\Phi$. To increase the SNR let us average, over neighbor reconstructed pixels, the correlation  $E_1E^*_0$  and the carrier intensity $|E_0|^2$:
\begin{eqnarray}\label{Eq_E1_E1}
|E_0|^2 &=& |E_{wo}|^2 ~J_0^2(\Phi) \simeq  |E_{wo}|^2 \Phi/2
\end{eqnarray}
By averaging we get:
\begin{eqnarray} \label{Eq_averaged E0E1}
  \langle E_1 E^*_0 \rangle_{x,y} &=& (1/N_{pix}) \sum_{x',y'} E_1(x',y') E^*_0(x',y')\\
    \nonumber &\simeq&  (\Phi(x,y)/2)~ (1/N_{pix}) \sum_{x',y'} |E_{wo}(x',y')|^2\\
\nonumber \langle |E_0|^2 \rangle_{x,y}&\simeq& (1/N_{pix})\sum_{x',y'} |E_{wo}(x',y')|^2
\end{eqnarray}
where $\sum_{x',y'}$ is the summation over the $N_{pix}$ pixels of the averaging zone located around the point of coordinate $x,y$. The vibration amplitude $\Phi$ is then
\begin{eqnarray} \label{Eq_Phi_E0EO_E1E0}
  \Phi(x,y) &=& 2 { \langle E_1 E^*_0 \rangle_{x,y}   }/{ \langle |E_0|^2 \rangle_{x,y} }
\end{eqnarray}
We get here $\Phi$  that gives both  the amplitude $z_{max}=\lambda |\Phi|/4\pi$ and the phase $\arg \Phi$ of the mechanical motion.
Note that it  is also possible to get $\Phi$ by calculating the ratio  $E_1/E_0\simeq \Phi$  as done by Verrier et. al \cite{verrier2013absolute}, but the ratio calculation is unstable for the points $x,y$ of the object where $E_{wo}$ is close to zero.

\begin{figure}
\begin{center}
  \includegraphics[width=8.0 cm]{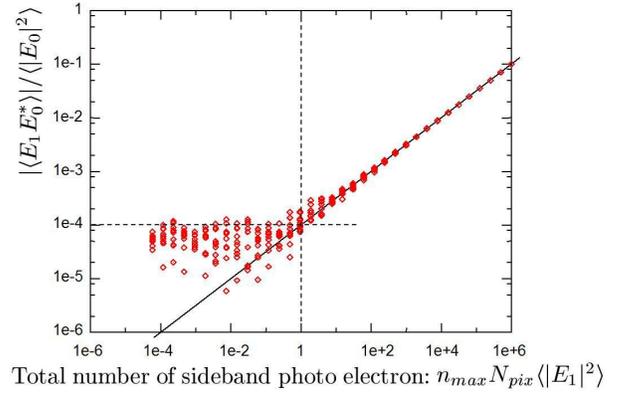}
 \caption{Ratio $|\langle E_1 E^*_0 \rangle |  / \langle |E_0|^2 \rangle$ calculated by  Monte Carlo by decreasing the sideband signal $|E_1|^2$.
Horizontal axis is the total sideband energy: $n_{max} N_{pix}\langle |E_1|^2\rangle $ in   photo electron Units.  Simulation is made with $n_{max}$=400, $N_{pix}=50^2$, $|E_{LO}|^2=10^4$ and $|E_{0}|^2=10^2$ photo electrons.} \label{Fig_fig_corr_th}
\end{center}
\end{figure}
To evaluate the limits of sensibility of the correlation + averaging method, we have calculated  by Monte Carlo the detection limit of $ |\Phi| $, for an ideal holographic detection  that is only limited by shot noise. The calculation is similar to one made by Lesaffre et al. \cite{lesaffre2013noise}.
For each pixel ($ x, y $),   each frame ($ n $)  and each sequence  ($m$=0 or $m$=1), we have calculated the ideal camera signal  $I'_n$ in the absence of shot noise. We have $ I'_n= | E_{LO} + j^n E'_ {0/1} |^2 $, where the factor $j^n$  accounts for the phase shift of  the local oscillator field $ E_{LO} $ with respect to object field $E_{0/1}$ for  frame $n$. To account for the roughness of the sample,   $ E'_{0} $ and $E'_1$ are taken  proportional to a Gaussian speckle $E'_{wo} (x, y)$ that is uncorrelated from one pixel to the next, but remains the same for all frames $n$ and all sequences $m$=0 or 1.
To account for shot noise, we have added to $I'_n$ a Gaussian random noise $ s $ of variance of $ \sqrt{I'_n} $, where $I'_n$ is expressed in electron photo electron Units. The noise  $ s $ is uncorrelated from one pixels $ x, y $ to another, from one frame $ n $ to another, and  from  one sequence $m $ to another.
By this way, we have obtained the  Monte Carlo frame signals $ I_n= I'_n  + s $ with whom we have performed the 4 phase demodulation with $n_{max}$ frames of Eq. (\ref{EQ_H_n_max}). We have then calculated the reconstructed signal $E_0 $ and $E_1$, the  correlations and intensities $E_1 E^*_0 $   and $ | E_0 |^2 $, and the means $ \langle E_1 E_0^* \rangle $ and $ \langle | E_0 |^2 \rangle $. We have then calculated the ratio $ {\langle E_1 E^*_0 \rangle} / {\langle | E_0 |^2 \rangle} $ that gives $ \Phi $ using Eq. (\ref {Eq_Phi_E0EO_E1E0}).

Figure \ref{Fig_fig_corr_th} gives the result of the Monte Carlo simulation for the ratio $| {\langle E_1 E^*_0 \rangle} |/ {\langle | E_0 |^2 \rangle} $. Each point correspond to simulation made with a  double sequence $ m $=0 and 1.
The simulation is performed by decreasing  the sideband averaged signal field intensity  $\langle | E_1|^2 \rangle $=1, 0.5, 0.25 ... photo electron per pixel and per frame. The other parameters of the simulation are
$n_{max}$= 400,  $N_{pix}$=$50^2$, $| E_{LO}|^2$=$10^4$ and $ \langle | E_0 |^2\rangle$=$10^2$ photo electrons. For each value of $\langle|E_1|^2\rangle$, 10 simulations are performed.
\begin{figure}
\begin{center}
  \includegraphics[width=8cm]{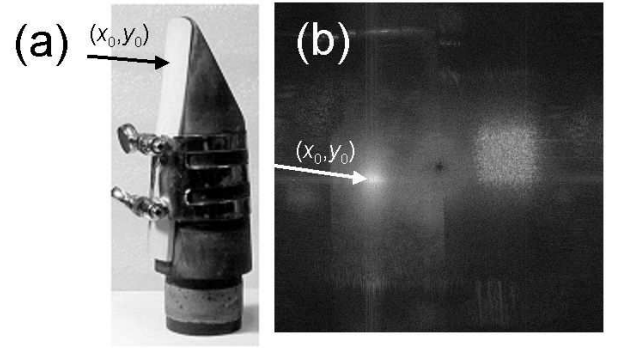}
 \caption{ (a) Clarinet reed with illumination beam focused in $x_0,y_0$. (b) Sideband $m=1$ reconstructed image of the vibrating reed. The display is made in arbitrary log scale for the field intensity $|E_1(x,y)|^2$.   } \label{Fig_fig_reed_1pt}
\end{center}
\end{figure}
As can be seen in Fig. \ref{Fig_fig_corr_th} the  ratio $| {\langle E_1 E^*_0 \rangle}| / {\langle | E_0 |^2 \rangle} $ decreases  with the sideband signal $ E_1$ proportionally with $ \sqrt{\langle|E_1|^2\rangle} $.  When $ E_1$ becomes very small, the ratio reaches a noise  floor related to shot noise.  To simplify the present discussion, the results  are displayed as a function of the total number of photos electrons on the sideband $ m =$1. The x-axis is thus $ n_{max} N_ {pix} \langle|E_1|^2\rangle$.
As seen of Fig.\ref{Fig_fig_corr_th} the noise floor is reached for $ n_{max} N_ {pix} \langle|E_1|^2\rangle = 1$. We have verified by making simulation not displayed on Fig.\ref{Fig_fig_corr_th} that this result do not depend on $ | E_{LO} |^2 $ and $ \langle | E_{0} |^2 \rangle$, provided that $ | E_{LO} |^2 \gg \langle | E_{0} |^2\rangle \gg \langle | E_{1} |^2 \rangle $. The noise floor corresponds thus
to a minimal ratio ${|\langle E_1 E^*_0 \rangle|} / {\langle | E_0 |^2 \rangle}= 1/ \sqrt{ n_{max} N_{pix} \langle| E_0 |^2\rangle}$ (i.e. to $10^{-4}$), and to a minimal detectable vibration amplitude $ \Phi = 2 / \sqrt { n_{max} N_{pix} \langle| E_0 |^2\rangle}$ (i.e. to $ 2~  10^{-4}$).
\begin{figure}[h]
\begin{center}
  \includegraphics[width=8 cm]{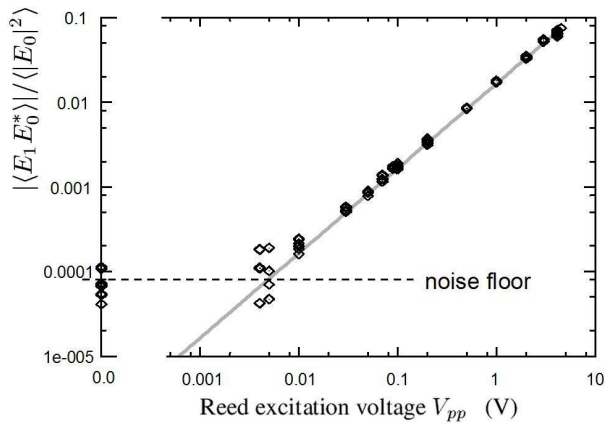}
 \caption{ Ratio $| \langle E_1 E^*_0 \rangle|/ \langle |E_0|^2 \rangle$  as function of the reed excitation voltage converted in vibration amplitude $z_{max}$ in nm Units. Measurements: dark grey points,  and theory   $J_1(|\Phi|)/J_0(|\Phi|)\simeq |\Phi|/2$ : light grey curve.   } \label{Fig_fig_lobna}
\end{center}
\end{figure}

We have tested the ability of the correlation + averaging method to measure low vibration amplitude in an experiment made with a vibrating clarinet reed excited at $\omega_A$=2 kHz with a loudspeaker.
In order to increase the sensitivity, the vibration amplitude is measured on a single point $(x_0,y_0)$ (see Fig.\ref{Fig_fig_reed_1pt} (a)). The  illumination has been focused on that point, and the calculations have been made with an averaging region centered on that point, whose size (radius 50 pixels for example) has been chosen to include the whole illumination zone. We have  reconstructed the field image of the reed at the carrier frequency (i.e. $E_0(x,y)$)  (see Fig.\ref{Fig_fig_reed_1pt} (b)),
and  verified on the holographic data that most of the energy $|E_0|^2$  is within the averaging zone  $x_0,y_0$.
Sequences of $n_{max}=128$ frames $I_n$ have been recorded for both carrier ($E_0$) and sideband ($E_1$), while the peak to peak voltage $V_{pp}$ of the   loudspeaker sinusoidal signal has been decreased.
We have then
    calculated $H$ by  Eq. (\ref{EQ_H_n_max}),
    reconstructed the fields images of the reed $E_0$ and $E_1$,
    and calculated the correlation  $E_1 E^*_0$, the intensity$|E_0|^2$, and the means $\langle E_1 E^*_0\rangle $ and $\langle|E_0|^2\rangle$.
 We have then calculated the ratio $ {|\langle E_1 E^*_0 \rangle|} / {\langle | E_0 |^2 \rangle} $. The latter is plotted on Fig.\ref{Fig_fig_lobna} as a function of the loudspeaker voltage $V_{pp}$ that is proportional to vibration amplitude $\Phi$. The measured points follow  $  {|\langle E_1 E^*_0 \rangle|} / {\langle | E_0 |^2 \rangle} \simeq |\Phi|/2 \propto V_{pp}$ down to a vibration amplitude   noise  floor of $|\Phi| \simeq 7 ~10^{-5}$ that corresponds to $z_{max}\simeq 10^{-2}$ nm i.e $\lambda/78000$. Note that the noise floor measure here is about $\times 20$ lower than in previous holographic experiments \cite{psota2012measurement,verrier2013absolute} and lower than the limit $\lambda/3500$ predicted by Ueda \cite{ueda1976signal}. Similar noise floor has been detected by holography by Bruno et al. \cite{bruno2014holographic}, but at much higher vibration frequency (40 kHz).

%
%

In future work, it would be interesting to explore  the limits of sensitivity of the correlation technique for low vibration amplitude. This can be done by using a laser with lower noise, by increasing the vibration frequency $\omega_A$ (and in all case by measuring the laser noise at $\omega_A$). One could also increase the illumination  power, and the number of frame of the coherent detection  ($n_{max}>128$). A better control of the vibration $\omega_A$ versus  camera  $\omega_{CCD}$ frequencies, and or a proper choice of the demodulation equation could be also important, in order to avoid leak detection of  $E_0$, when detection is tuned on $E_1$.

The Authors wishes to acknowledge ANR-ICLM (ANR-2011-BS04-017-04) for financial support, and M. Lesaffre for fruitful discussion and helpful comments.


\begin{thebibliography}{0}%
\makeatletter
\providecommand \@ifxundefined [1]{%
 \@ifx{#1\undefined}
}%
\providecommand \@ifnum [1]{%
 \ifnum #1\expandafter \@firstoftwo
 \else \expandafter \@secondoftwo
 \fi
}%
\providecommand \@ifx [1]{%
 \ifx #1\expandafter \@firstoftwo
 \else \expandafter \@secondoftwo
 \fi
}%
\providecommand \natexlab [1]{#1}%
\providecommand \enquote  [1]{``#1''}%
\providecommand \bibnamefont  [1]{#1}%
\providecommand \bibfnamefont [1]{#1}%
\providecommand \citenamefont [1]{#1}%
\providecommand \href@noop [0]{\@secondoftwo}%
\providecommand \href [0]{\begingroup \@sanitize@url \@href}%
\providecommand \@href[1]{\@@startlink{#1}\@@href}%
\providecommand \@@href[1]{\endgroup#1\@@endlink}%
\providecommand \@sanitize@url [0]{\catcode `\\12\catcode `\$12\catcode
  `\&12\catcode `\#12\catcode `\^12\catcode `\_12\catcode `\%12\relax}%
\providecommand \@@startlink[1]{}%
\providecommand \@@endlink[0]{}%
\providecommand \url  [0]{\begingroup\@sanitize@url \@url }%
\providecommand \@url [1]{\endgroup\@href {#1}{\urlprefix }}%
\providecommand \urlprefix  [0]{URL }%
\providecommand \Eprint [0]{\href }%
\providecommand \doibase [0]{http://dx.doi.org/}%
\providecommand \selectlanguage [0]{\@gobble}%
\providecommand \bibinfo  [0]{\@secondoftwo}%
\providecommand \bibfield  [0]{\@secondoftwo}%
\providecommand \translation [1]{[#1]}%
\providecommand \BibitemOpen [0]{}%
\providecommand \bibitemStop [0]{}%
\providecommand \bibitemNoStop [0]{.\EOS\space}%
\providecommand \EOS [0]{\spacefactor3000\relax}%
\providecommand \BibitemShut  [1]{\csname bibitem#1\endcsname}%
\let\auto@bib@innerbib\@empty
\end{thebibliography}%


\begin{thebibliography}{10}
\bibitem{pedrini1995digital}
G.~Pedrini, Y.~Zou, and H.~Tiziani, ``Digital double-pulsed holographic
  interferometry for vibration analysis,'' J. Mod. Opt. {\bf 42},
  367--374 (1995).

\bibitem{pedrini1997digital}
G.~Pedrini, H.~J. Tiziani, and Y.~Zou, ``Digital double
  pulse-TV-holography,'' Opt. Las. Eng. {\bf 26}, 199--219
  (1997).

\bibitem{pedrini1998transient}
G.~Pedrini, P.~Froening, H.~Fessler, and H.~Tiziani, ``Transient
  vibration measurements using multi-pulse digital holography,'' Opt. Las.
  Technol. {\bf 29}, 505--511 (1998).

\bibitem{pedrini2006high}
G.~Pedrini, W.~Osten, and M.~E. Gusev, ``High-speed digital holographic
  interferometry for vibration measurement,'' Appl. Opt. {\bf 45},
  3456--3462 (2006).

\bibitem{fu2007vibration}
Y.~Fu, G.~Pedrini, and W.~Osten, ``Vibration measurement by temporal
  Fourier analyses of a digital hologram sequence,'' Appl. Opt. {\bf 46},
  5719--5727 (2007).

\bibitem{powell1965interferometric}
R.~L. Powell and K.~A. Stetson, ``Interferometric vibration analysis by
  wavefront reconstruction,'' J. Opt. Soc. A {\bf 55}, 1593--1597 (1965).

\bibitem{picart2003time}
P.~Picart, J.~Leval, D.~Mounier, and S.~Gougeon, ``Time-averaged digital
  holography,'' Opt. Lett. {\bf 28}, 1900--1902 (2003).

\bibitem{picart2005some}
P.~Picart, J.~Leval, D.~Mounier, and S.~Gougeon, ``Some opportunities
  for vibration analysis with time averaging in digital Fresnel holography,''
  Appl. Opt. {\bf 44}, 337--343 (2005).

\bibitem{le2000numerical}
F.~Le~Clerc, L.~Collot, and M.~Gross, ``Numerical heterodyne holography
  with two-dimensional photodetector arrays,'' Opt. Lett. {\bf 25}, 716--718
  (2000).

\bibitem{le2001synthetic}
F.~Le~Clerc, M.~Gross, and L.~Collot, ``Synthetic-aperture experiment in
  the visible with on-axis digital heterodyne holography,'' Opt. Lett. {\bf
  26}, 1550--1552 (2001).

\bibitem{ueda1976signal}
M.~Ueda, S.~Miida, and T.~Sato, ``Signal-to-noise ratio and smallest
  detectable vibration amplitude in frequency-translated holography: an
  analysis,'' Appl. Opt. {\bf 15}, 2690--2694 (1976).

\bibitem{psota2012measurement}
P.~Psota, V.~Ledl, R.~Dolecek, J.~Erhart, and V.~Kopecky, ``Measurement
  of piezoelectric transformer vibrations by digital holography,'' IEEE T. Ultrason. Ferr. {\bf 59},
  1962--1968 (2012).

\bibitem{psota2012comparison}
P.~Psota, V.~L{\'e}dl, R.~Dole{\v{c}}ek, J.~V{\'a}clav{\'\i}k, and
  M.~{\v{S}}ulc, ``Comparison of digital holographic method for very
  small amplitudes measurement with single point laser interferometer and laser
  Doppler vibrometer,'' in {\em Digital Holography and Three-Dimensional
  Imaging\/}  pp. DSu5B--3 (2012).

\bibitem{verrier2013absolute}
N.~Verrier and M.~Atlan, ``Absolute measurement of small-amplitude
  vibrations by time-averaged heterodyne holography with a dual local
  oscillator,'' Opt. Lett. {\bf 38}, 739--741 (2013).

\bibitem{bruno2014phase}
F.~Bruno, J.-B. Laudereau, M.~Lesaffre, N.~Verrier, and M.~Atlan,
  ``Phase-sensitive narrowband heterodyne holography,'' Appl. Opt.
  {\bf 53}, 1252--1257 (2014).

\bibitem{bruno2014holographic}
F.~Bruno, J.~Laurent, D.~Royer, and M.~Atlan, ``Holographic imaging of
  surface acoustic waves,'' Appl. Phys. Lett. {\bf 104}, 083\,504 (2014).

\bibitem{schnars1994direct}
U.~Schnars and W.~Juptner, ``Direct recording of holograms by a CCD
  target and numerical reconstruction,'' Appl. Opt. {\bf 33}, 179--181
  (1994).

\bibitem{lesaffre2013noise}
M.~Lesaffre, N.~Verrier, and M.~Gross, ``Noise and signal scaling
  factors in digital holography in weak illumination: relationship with shot
  noise,'' Appl. Opt. {\bf 52}, A81--A91 (2013).

\end{thebibliography}
\end{document}